\documentclass[english]{article}
\usepackage[T1]{fontenc}
\usepackage[latin9]{inputenc}
\usepackage{amsmath}
\usepackage{babel}
\begin{document}
\title{On the curvature extrema\\
of special cubic Bézier curves}
\author{Kenjiro T.~Miura, Péter Salvi}
\maketitle

\section{Introduction}

In this document we are going to prove that a cubic Bézier curve
\begin{equation}
C(t)=\sum_{i=0}^{3}P_{i}\binom{3}{i}t^{i}(1-t)^{3-i}
\end{equation}
with the special control point configuration
\begin{align}
P_{0} & =Q_{0}, & P_{1} & =(1-a)Q_{0}+aQ_{1}, & P_{2} & =aQ_{1}+(1-a)Q_{2}, & P_{3} & =Q_{2}
\end{align}
has at most one local extremum of curvature when $t\in(0,1)$ and
$a\in(\frac{2}{3},1]$.

\section{Special cases}

We need to treat two special cases first. When $Q_{0}=Q_{2}$, the
curve degenerates to a line segment, which has a kink (a point with
infinite curvature) at $t=\frac{1}{2}$, so it clearly has exactly
one local curvature extremum.

With this handled, let us set, without loss of generality,
\begin{align}
Q_{0} & =(-1,0), & Q_{1} & =(b,h), & Q_{2} & =(1,0),\label{eq:cp}
\end{align}
where $b,h\geq0$. The second special case is when $h=0$. Once again,
the curve degenerates to a line segment. When $b\in(-1,1)$, its curvature
is always $0$, otherwise it has a single kink.

In the following we will assume $h>0$.

\section{Curvature extrema}

The signed curvature of a planar polynomial curve $(x(t),y(t))$ is
given~\cite{doCarmo} as
\begin{equation}
\kappa=\frac{x'y''-x''y'}{\left(x'^{2}+y'^{2}\right)^{\frac{3}{2}}},
\end{equation}
where $x'$ is the first derivative of $x$ with respect to $t$ etc.
The derivative of its square is 
\begin{align}
(\kappa^{2})'=2\kappa\kappa' & =\frac{2\left(x'y''-x''y'\right)\left(x'y'''-x'''y\right)\left(x'^{2}+y'^{2}\right)^{3}}{\left(x'^{2}+y'^{2}\right)^{6}}\nonumber \\
 & -\frac{\left(x'y''-x''y'\right)^{2}3\left(x'^{2}+y'^{2}\right)^{2}\left(2x'x''+2y'y''\right)}{\left(x'^{2}+y'^{2}\right)^{6}}
\end{align}
so
\begin{align}
\left(x'^{2}+y'^{2}\right)^{4}\kappa\kappa' & =\left(x'y''-x''y'\right)\left[\left(x'y'''-x'''y\right)\left(x'^{2}+y'^{2}\right)\right.\nonumber \\
 & \quad\qquad\qquad\qquad\ \left.-3\left(x'y''-x''y'\right)\left(x'x''+y'y''\right)\right].
\end{align}
Here $x'y''-x''y'=0$ means that the curvature is equal to $0$, and
it corresponds to inflection points. Consequently,
\begin{equation}
\left(x'y'''-x'''y\right)\left(x'^{2}+y'^{2}\right)-3\left(x'y''-x''y'\right)\left(x'x''+y'y''\right)=0\label{eq:dk}
\end{equation}
corresponds to curvature extrema.

\section{Main proof}

Let $N(t,a)$ denote the left-hand side of Eq.~(\ref{eq:dk}), applied
to our curve, with control points given as in (\ref{eq:cp}). We need
to show that $N(t,a)$ has at most one 0-crossing when $t\in(0,1)$
and $a\in(\frac{2}{3},1]$.

First note that
\begin{equation}
N(0,a)=-324a^{2}h\left[(1+b)(12+3a^{2}(5+b)-4a(7+b))+a(-4+3a)h^{2}\right],
\end{equation}
which is positive for $a=\frac{2}{3}$. Let us denote the expression
in brackets with $f_{0}(a)$. We have $f_{0}(\frac{2}{3})<0$, $f_{0}(1)<0$
and
\begin{equation}
\frac{df_{0}(a)}{da}=6(b^{2}+6b+5+h^{2})a-4(b^{2}+8b+7+h^{2}),
\end{equation}
from which we can see that $df_{0}(a)/da$ is negative at $a=\frac{2}{3}$,
and increases with $a$. We can conclude that as $a$ goes from $\frac{2}{3}$
to $1$, $f_{0}$ first decreases, and then it may increase, but since
$f_{0}(1)<0$, it always remains negative, so
\begin{equation}
N(0,a)>0.
\end{equation}

The derivative of $N$ with respect to $t$ is given as
\begin{equation}
\frac{\partial N(t,a)}{\partial t}=1296ah\cdot f_{1}(t,a)\cdot f(t,a),
\end{equation}
where
\begin{align}
f_{1}(t,a) & =2t^{2}-2t+1-\left(3t^{2}-3t+1\right)a,\\
f(t,a) & =-3a^{2}(b^{2}+b(10-20t)+h^{2}+60(t-1)t+13)\nonumber \\
 & +4a(5b(1-2t)+60(t-1)t+11)+80(1-t)t-12.
\end{align}
It is easy to see that $f_{1}(t,a)\geq0$, since its derivative with
respect to $a$ is negative, and $f_{1}(t,1)$ is positive. Consequently,
the signs of $\partial N(t,a)/\partial t$ and $f(t,a)$ are the same.

As for $f(t,a)$, note that it is a quadratic function of $t$, and
the coefficient of $t^{2}$ is $-20(9a^{2}-12a+4)$, which is negative
for $a\in(\frac{2}{3},1]$. We can also prove
\begin{equation}
f(0,a)<0,\label{eq:f0}
\end{equation}
by ascertaining that $f(0,\frac{2}{3})<0$, $\partial f(0,\frac{2}{3})/\partial a<0$
and $\partial^{2}f(0,a)/\partial a^{2}<0$, as this means that $\partial f(0,a)/\partial a$
decreases as $a$ goes from $\frac{2}{3}$ to $1$, starting from
a negative value, and thus $f(0,a)$ itself is always negative, as
well.

\subsection{Case I: $b\protect\leq3-\frac{2}{a}$}

The maximum of $f(t,a)$ for a fixed $a$ value is found by solving
$\partial f(t,a)/\partial t=0$, giving
\begin{equation}
t_{0}=\frac{ab+3a-2}{2(3a-2)}.
\end{equation}
When $b\leq3-\frac{2}{a}$, this will be in the $[0,1]$ interval,
and
\begin{align}
f(t_{0},a) & =8-16a+6a^{2}-3a^{2}h^{2}+2a^{2}b^{2}\\
 & <8-16a+6a^{2}-3a^{2}h^{2}+2a^{2}\left(3-\frac{2}{a}\right)^{2}\\
 & =(24-3h^{2})a^{2}-40a+16=:f_{3}(a).
\end{align}

Since $f_{3}$ is quadratic in $a$, and also $f_{3}(\frac{2}{3})<0$,
$f_{3}(1)<0$ and $df_{3}(\frac{2}{3})/da<0$, we can see that as
$a$ goes from $\frac{2}{3}$ to $1$, the value of $f_{3}$ decreases,
starting from a negative value, and while it may start to increase,
it remains negative.

In summary, we have shown that in this case the maximum of $f$ is
negative, so $\partial N(t,a)/\partial t<0$, i.e., the curvature
decreases monotonically.

\subsection{Case II: $b>3-\frac{2}{a}$}

In this case $f$ takes its maximum over the $[0,1]$ interval at
$t=1$. We can also state the following (these will be proved below):
\begin{align}
N(1,a) & <0\text{ when }f(1,a)>0,\label{eq:n1}\\
\frac{\partial f(0,a)}{\partial t} & >0,\label{eq:df0}\\
\frac{\partial^{2}f(t,a)}{\partial t^{2}} & <0.\label{eq:d2f}
\end{align}
Equation~(\ref{eq:d2f}) shows that $\partial f(t,a)/\partial t$
decreases monotonically as $t$ goes from $0$ to $1$, while by Eq.~(\ref{eq:df0})
it starts from a positive value, so it may have at most one 0-crossing.
Consequently $f$ first increases, starting from a negative value
(Eq.~\ref{eq:f0}), and then it may decrease.

When $f(1,a)<0$, since this is its maximal value, it means that $f$
is always negative, so the curvature decreases monotonically.

Otherwise $f$ has exactly one 0-crossing, so $N(t,a)$ first decreases,
starting from a positive value, and then increases, as $t$ goes from
$0$ to $1$. Since in this case $N(1,a)<0$ (Eq.~\ref{eq:n1}),
there is exactly one curvature extremum. 

\subsubsection{Proof of Eq.~(\ref{eq:n1})}

Since
\begin{equation}
f(1,a)=-3a^{2}\left(\frac{15a-10}{3a}-b\right)^{2}-3a^{2}h^{2}+36\left(a-\frac{2}{3}\right)\left(a-\frac{8}{9}\right),
\end{equation}
it can be seen that $f(1,a)$ can only be positive when $a>\frac{8}{9}$.
Under these constraints it holds that $15a-10>3a$, so
\begin{equation}
f(1,a)|_{b<1}<f(1,a)|_{b=1}<0.
\end{equation}
Consequently, the assumption $f(1,a)>0$ implies $b>1$.

Restructuring
\begin{align}
N(1,a) & =-324a^{2}h\left[12(b-1)+4a(7+(b-8)b+h^{2})-3a^{2}(5+(b-6)b+h^{2})\right]\\
 & =-324a^{3}(4-3a)h\left[\left(b-\frac{-9a^{2}+16a-6}{(4-3a)a}\right)^{2}-\frac{36(a-1)^{4}}{(4-3a)^{2}a^{2}}+h^{2}\right]
\end{align}
shows that $N(1,a)$ is negative when $(b,h)$ is outside the circle
with center $\left(\frac{-9a^{2}+16a-6}{(4-3a)a},0\right)$ and radius
$\frac{6(a-1)^{2}}{(4-3a)a}$. But $(b,h)$ will be outside for any
$b>1$, which proves Eq.~(\ref{eq:n1}).

\subsubsection{Proof of Eq.~(\ref{eq:df0})}

We have
\begin{equation}
\frac{\partial f(0,a)}{\partial t}=20\left[a(3a-2)b+3a(3a-4)+4\right],
\end{equation}
and $3a(3a-4)+4>0$ for any $a\in(\frac{2}{3},1]$, Eq.~(\ref{eq:df0})
is satisfied.

\subsubsection{Proof of Eq.~(\ref{eq:d2f})}

Derivating twice,
\begin{equation}
\frac{\partial^{2}f(t,a)}{\partial t^{2}}=40(12a-9a^{2}-4),
\end{equation}
which is negative for all $a\in(\frac{2}{3},1]$.

\end{document}